\documentclass[pdflatex,sn-nature]{sn-jnl}

\usepackage{graphicx}%
\usepackage{multirow}%
\usepackage{amsmath,amssymb,amsfonts}%
\usepackage{amsthm}%
\usepackage{mathrsfs}%
\usepackage{lmodern}
\usepackage[title]{appendix}%
\usepackage{xcolor}%
\usepackage{textcomp}%
\usepackage{manyfoot}%
\usepackage{booktabs}%
\usepackage{algorithm}%
\usepackage{algorithmicx}%
\usepackage{algpseudocode}%
\usepackage{listings}%
\usepackage{array}
\usepackage{multirow}
\usepackage{verbatim}
\usepackage{xcolor}

\usepackage{adjustbox}
\usepackage{url}
\newcommand{\ours}{ObsCast}
\usepackage{svg}
\usepackage{hyperref}
\usepackage{booktabs}
\usepackage{tabularx}
\usepackage{array}
\usepackage{newfloat}
\setcitestyle{open={},close={},comma,sort&compress,super}

\DeclareFloatingEnvironment[
    name={Extended Data Fig.}
]{extdatafig}

\begin{document}
\title[]{Skillful high-resolution weather forecasting independent of physical models}
% Unified weather analysis and forecasting learned directly from observations
% \ours{}: 

\author[1]{\fnm{Pengcheng} \sur{Zhao}}
\equalcont{These authors contributed equally to this work.}

\author[1]{\fnm{Siqi} \sur{Xiang}}
\equalcont{These authors contributed equally to this work.}

\author[1]{\fnm{Weixin} \sur{Jin}}
\equalcont{These authors contributed equally to this work.}

\author[1]{\fnm{Zekun} \sur{Ni}}

\author[1]{\fnm{Jiang} \sur{Bian}}

\author[1]{\fnm{Zuliang} \sur{Fang}}

\author[1]{\fnm{Hongyu} \sur{Sun}}

\author[1]{\fnm{Bin} \sur{Zhang}}

\author[2,3]{\fnm{Richard} \sur{E. Turner}}

\author[1]{\fnm{Jonathan} \sur{Weyn}}

\author*[1]{\fnm{Haiyu} \sur{Dong}}
\email{haiyu.dong@microsoft.com}

\author*[1]{\fnm{Kit} \sur{Thambiratnam}}
\email{kitth@microsoft.com}

\author*[1]{\fnm{Qi} \sur{Zhang}}
\email{zhang.qi@microsoft.com}

\affil[1]{\orgname{Microsoft Corporation}}

\affil[2]{\orgname{University of Cambridge}, \orgaddress{\city{Cambridge}, \country{United Kingdom}}}

\affil[3]{\orgname{The Alan Turing Institute}, \orgaddress{\city{London}, \country{United Kingdom}}}

\abstract{
    Accurate and timely weather forecasts are critical for high-impact decisions in modern society.
    Machine-learning-based weather prediction is emerging as an alternative for producing initial conditions, forecasts, and even both in end-to-end systems. These methods deliver predictions faster and often with higher skill than traditional numerical weather prediction (NWP).
    However, even end-to-end models typically rely on NWP-generated reanalyses for supervision, thereby inheriting the biases and resolution limitations of those NWPs, and limiting adaptation to settings where suitable reanalysis products are unavailable, infrequently updated, or expensive to produce.
    Here we introduce \ours{}, a regional system that generates both analysis and predictions, without using any NWP-derived data in either training or inference, while still achieving state-of-the-art performance in short-term high-resolution regional modeling.
    Over the contiguous United States and Europe, \ours{} outperforms operational NWP for near-surface variables through 18 h and produces skillful precipitation forecasts.
    It provides a simpler and more adaptable route to build and refine regional forecasting services directly from local observations, without the need to develop complex and costly traditional forecasting pipelines.

    % Accurate and timely weather forecasts are critical for high-impact decisions in modern society.
    % Machine-learning-based weather prediction (MLWP) is beginning to replace data assimilation (DA), forecasting, and even both in end-to-end systems. These methods deliver predictions more rapidly and often with higher skill than traditional numerical weather prediction (NWP).
    % However, these systems are trained using reanalyses, inheriting the biases and resolution limitations of the underlying NWPs.
    % Our model, \ours{}, circumvents these issues by using no NWP-derived data throughout both training and operation, while still achieving state-of-the-art accuracy in short-term high-resolution regional modeling. 
    % Over CONUS and Europe, \ours{} outperforms operational NWP for near-surface temperature, wind, humidity, and pressure forecasts [by X\%] through 18 h and yields skillful precipitation forecasts.
    % ObsCast provides a practical route to build and iterate regional services directly from local observations, without the need to develop and NWP-DA pipeline.
}

\maketitle

Since the mid-20th century, numerical weather prediction (NWP) has served as the foundation of weather forecasting~\cite{bauer2015quiet}. These models predict the evolution of the atmosphere from current conditions mainly by numerically solving the governing partial differential equations of atmospheric dynamics and thermodynamics. In parallel, NWP has driven the development of modern data assimilation (DA) systems, which use model dynamics to blend a forecast background with heterogeneous observations into gridded estimates of the atmospheric state~\cite{kalnay2003atmospheric, geer2018all}. However, this DA-NWP workflow remains computationally intensive and is not designed to directly learn from the growing archives of historical data~\cite{reichstein2019deep, ben2024rise}.

In recent years, a profound transformation in this area has been the use of machine-learning-based weather prediction (MLWP)~\cite{bi2023accurate, chen2023fuxi, lam2023learning, kochkov2024neural, lang2024aifs, price2025probabilistic, bodnar2025foundation}. The maturation of DA has produced multi-decadal reanalyses, which serve as accessible, continuous, and physically consistent training datasets~\cite{parker2016reanalyses, hersbach2020era5}. Consequently, many of the first MLWP models were trained on reanalysis fields. These models have achieved higher forecast skill than NWP while substantially reducing computational costs. However, training on reanalyses means that models learn from atmospheric states produced by NWP-based DA systems, and therefore reflect the biases and errors of those systems~\cite{parker2016reanalyses, ben2024rise}. Moreover, in operational settings, they still rely on initial conditions provided by the conventional workflow, meaning that they remain dependent on DA for initialization, with both latency and forecast skill constrained by that system.

To further reduce reliance on NWP, recent work has explored end-to-end learning~\cite{chen2023towards, alexe2024graphdop, allen2025end, sun2025data}, a paradigm emerging across scientific fields that replaces empirical, rule-based components with fully learnable pipelines~\cite{litjens2017survey, choudhary2022recent}. In weather forecasting, this approach integrates DA modules to ingest raw observations directly, thereby replacing the entire NWP pipeline~\cite{schultz2021can}. However, this transition remains incomplete, as most systems still rely on NWP-derived reanalyses during training~\cite{chen2023towards, allen2025end, sun2025data}. A distinct line of research explores training and initializing forecasting models exclusively from raw observations, allowing models to learn atmospheric evolution from observations and discover physical patterns that are imperfectly represented in current NWP systems~\cite{mcnally2024data, alexe2024graphdop}. However, unlike reanalysis-based AI models that have already surpassed operational benchmarks, these end-to-end observation-driven models currently exhibit a significant performance gap compared to operational NWP systems. Furthermore, they are implemented at relatively coarse spatial resolutions, typically around 100 km~\cite{alexe2024graphdop, allen2025end}. Together, these limitations highlight the difficulty of achieving high forecast skill and resolution without access to the reanalyses.

In this paper, we introduce \ours{}, which, to our knowledge, is the first regional framework that learns only from observations, while achieving competitive forecast performance with operational NWP systems at high spatial resolution. \ours{} is a regional model that produces gridded analyses and forecasts at 0.05° resolution for lead times up to 18 hours over a target domain. It formulates regional forecasting as two coupled learning problems: an analysis model learns dense gridded fields from sparse station-based supervision, and a forecast model learns to predict their temporal evolution autoregressively. We first train \ours{} over the data-rich continental United States (CONUS). It achieves consistently lower root mean square error (RMSE) than operational NWP systems across 1–18 h lead times for key near-surface variables. Precipitation verification further shows a clear early-lead advantage, with markedly stronger skill in the initial forecast hours. We then demonstrate its adaptability by extending the framework to the European domain, showing similar gains relative to NWP benchmarks. By avoiding reliance on the assumptions embedded in NWP systems, \ours{} may enable further performance gains. This paradigm enables organizations operating at regional scales to obtain and keep improving high-resolution, operationally ready analyses and forecasts at lower cost, without requiring investment in local NWP infrastructure.

\section{\ours{}}\label{overview}

\begin{figure}[htbp]
    \centering
\includegraphics[width=1\linewidth]{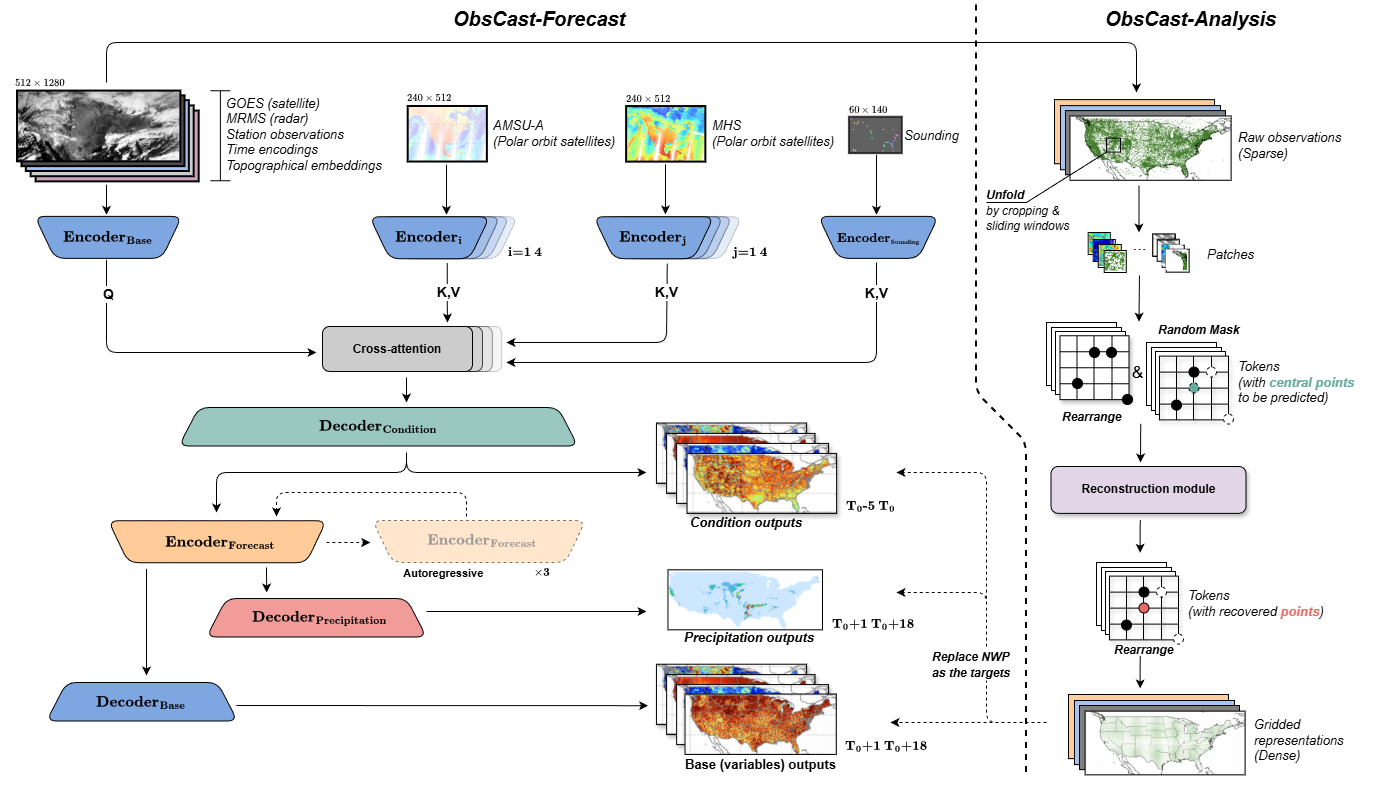}
    \caption{\textbf{Overview of the \ours{} framework.} The system comprises two components: an analysis model (right) that reconstructs gridded atmospheric fields from raw observations, and a forecast model (left) that predicts their temporal evolution. During training, the analysis model is supervised using surface station observations as labels and is subsequently used to generate training labels for the forecast model, which learns to produce autoregressive predictions directly from raw observations. Both training and inference are entirely independent of NWP; during inference, only the forecast model is executed.
}
    \label{fig:architecture}
\end{figure}

\ours{} is an end-to-end deep learning system that delivers multivariate hourly analyses and forecasts over a configurable target region at 0.05° spatial resolution. In this study, it covers 2-m temperature (T2M), 10-m wind speed (WS10M), 2-m specific humidity (Q2M), 2-m dew-point temperature (TD2M), and total precipitation (TP), and all results are restricted to land grid points. The framework comprises two components: an analysis model that converts raw observations into high-resolution gridded analysis fields, and a forecast model trained with these analysis fields as labels to produce hourly predictions up to an 18 h lead time (Figure~\ref{fig:architecture}).

First, the analysis model constructs regional gridded analysis fields from raw observations over the target region. The inputs include gridded observations from geostationary satellites and radar, together with irregularly spaced near-surface weather station measurements, similar to other DA systems~\cite{manshausen2025generative, xiang2025adaf, fan2026physically}. During training, supervisions are restricted to grid cells collocated with stations. For each target grid cell for a specific hour, the model ingests a local spatiotemporal window spanning one hour before and after the target time. At inference, the same network is applied at every grid cell to generate a dense analysis field.

Once the analysis model is trained, it is used to generate a historical analysis dataset that serves as training labels for the forecast model. The forecast model is also designed to ingest heterogeneous observations from multiple sources. Unlike the analysis model, the forecast model conditions on observations from the preceding 6 hours. In addition to the observations within the target region, polar orbiting satellite products and radiosonde observations are introduced over a larger spatial extent, providing broader upper-level dynamical context and lateral boundary information that supports longer forecast lead times~\cite{warner1997tutorial}. The model is rolled out autoregressively in 6-hour increments, so that three steps produce forecasts out to 18 hours. The conditioning and rollout steps are trained end-to-end under a unified loss function, constraining the physically supervised channels of the initial-state representation (see~\nameref{sec:methods}) while jointly optimizing the autoregressive forecasting.

Taken together, this two-stage architecture provides an end-to-end analysis and forecasting system that operates without reliance on NWP. The analysis model produces high-resolution, spatially complete estimates of the atmospheric state that are anchored directly to the raw observations. These analyses in turn enables the forecast model to transform heterogeneous, multi-source observations into high-resolution gridded predictions in both training and operational deployment.

\section{Results}

\subsection{Evaluation of \ours{}-Analysis}\label{anl}

\ours{} is designed as a regional model. We trained paired analysis and forecasting components over the contiguous United States (CONUS). The analysis component produces gridded estimates of the atmospheric state at each time step. We benchmark four near-surface variables against the Real-Time Mesoscale Analysis (RTMA), a 2.5-km, hourly gridded analysis over CONUS produced operationally by NOAA’s National Centers for Environmental Prediction (NCEP)~\cite{de2011real}. Because RTMA is designed to closely fit observations rather than to be used as initial conditions for a forecast model, it provides a suitable reference for evaluating \ours{}-Analysis. For precipitation, we compare hourly accumulations with the Multi-Radar Multi-Sensor (MRMS) quantitative precipitation estimation (QPE), which integrates data from multiple radar networks and additional sources to produce high-resolution precipitation estimates~\cite{zhang2016multi}. To quantify how well the analysis captures these conditions, we treat station observations as ground truth and interpolate all gridded products to station locations (see~\nameref{sec:methods}). In addition to a temporal split into training, validation, and test periods, we reserve a subset of stations as a held-out test set that are excluded from training and validation and used solely for testing.

\begin{figure}[htbp]
    \centering
\includegraphics[width=1\linewidth]{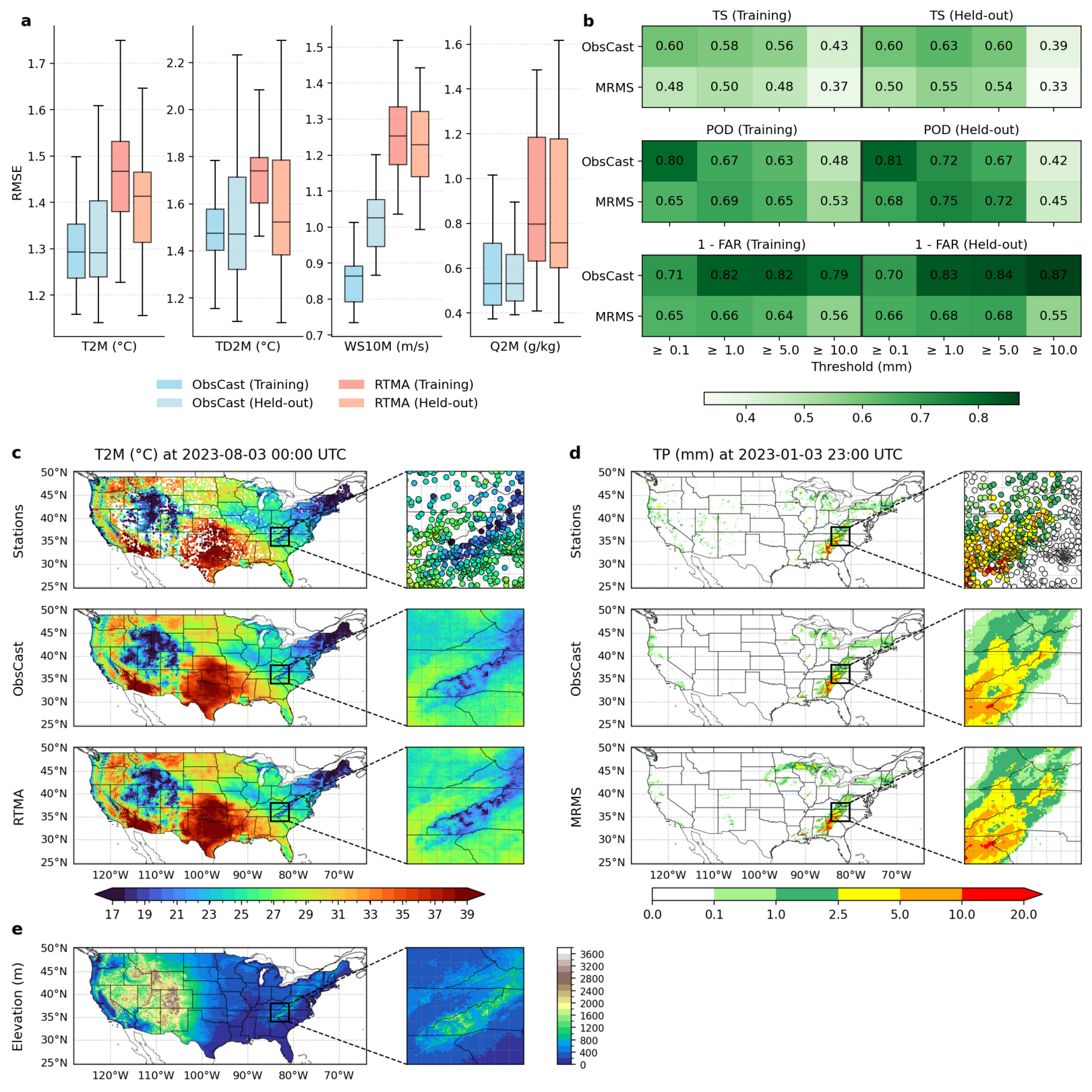}
    \caption{\textbf{Evaluation of \ours{}-Analysis over CONUS.} \textbf{a}, Station-wise RMSE of near-surface variables for \ours{}-Analysis and the RTMA baseline, evaluated against surface station observations. Results are shown separately for stations used during training and for held-out stations, demonstrating generalization to unseen locations. \textbf{b}, Skill of hourly accumulated precipitation analysis relative to gauge observations, quantified using TS, POD, and 1$-$FAR at multiple intensity thresholds. Metrics are shown for both training and held-out stations, with MRMS used as the reference baseline. \textbf{c}, Example of 2-m air temperature at 00:00 UTC on 3 August 2023, shown as station observations and corresponding gridded values from \ours{}-Analysis and RTMA. \textbf{d}, Example of hourly accumulated precipitation at 01:00 UTC on 3 January 2023, shown as station observations and corresponding gridded values from \ours{}-Analysis and MRMS. \textbf{e}, Elevation distribution of surface stations over CONUS land. 
}
    \label{fig:anl-metrics}
\end{figure}

Figure~\ref{fig:anl-metrics}a shows station-wise RMSE for different near-surface variables. \ours{}-Analysis attains lower RMSE than RTMA for all variables, with improvements evident in both the median and, for most variables, a reduced spread across stations. On stations used for training, RMSE of 2-m temperature, 2-m dew-point temperature, 10-m wind speed, and 2-m specific humidity is reduced by 11.2\%, 12.9\%, 31.8\%, and 34.5\% respectively relative to RTMA. Importantly, the system maintains its advantage over RTMA at held-out stations, demonstrating robust generalization to previously unseen locations. For these held-out stations, the overall RMSE reductions relative to RTMA are 5.9\%, 2.9\%, 17.5\%, and 31.5\% for the same variables. This indicates that the model learns the underlying physics within the input window and produces accurate analyses at the target center point, rather than memorizing specific locations. Consequently, it supports the reliability of the analyses at locations that cannot be evaluated directly because station observations are unavailable.

\ours{} also produces skillful analyses of hourly precipitation accumulation. We evaluate hourly accumulation using three standard threshold-based metrics: Threat Score (TS), Probability of Detection (POD), and False Alarm Ratio (FAR), at thresholds of 0.1, 1.0, 5.0 and 10.0 mm/h. Figure~\ref{fig:anl-metrics}b compares \ours{}-Analysis and MRMS against gauge-based station observations (for consistency, FAR is shown as $1-FAR$). Performance is highly consistent between stations used for training and held-out stations. \ours{}-Analysis outperforms MRMS on most metrics, except that POD is lower at the 1.0, 5.0, and 10.0 mm/h thresholds. Nevertheless, TS, which combines the effects of POD and FAR, is 16.0–25.0\% higher than MRMS on training stations and 11.1–20.0\% higher on held-out stations. These results suggest that \ours{}-Analysis uses precipitation-related signatures from satellites, radar, and nearby stations to reconstruct precipitation patterns.

Figure~\ref{fig:anl-metrics}c-d shows examples of \ours{}-Analysis fields for 2-m temperature and precipitation. \ours{}-Analysis reproduces the large-scale temperature patterns and topography-related structure that are consistent with both station observations and RTMA. In the enlarged view, temperature variations along the southern Appalachian Mountains align with elevation (Figure~\ref{fig:anl-metrics}e), indicating that \ours{}-Analysis captures fine-scale topographic modulation. Figure~\ref{fig:anl-metrics}d shows a precipitation event over the eastern CONUS. \ours{}-Analysis closely reproduces the observed precipitation structure, including the spatial extent and intensity gradient of the frontal rainband. \ours{}-Analysis exhibits smoother and more coherent spatial transitions in the peripheral regions of the rainband compared to MRMS, which is consistent with its higher POD at low thresholds in Figure~\ref{fig:anl-metrics}b. These examples suggest that \ours{}-Analysis can reconstruct high-resolution gridded fields from raw observations.

\subsection{Evaluation of \ours{}-Forecast}\label{results}

Next, we used the CONUS-trained analysis model to generate a historical dataset, which we then used to train a forecast model for CONUS. The forecast model takes raw observations as input and is trained to predict \ours{}-Analysis. The temporal split into training, validation, and test periods, as well as the choice of held-out stations, was kept identical to the analysis model setup to prevent information leakage between training and evaluation. We compare the forecast skill of \ours{} with deterministic forecasts from the High-Resolution Rapid Refresh (HRRR)~\cite{dowell2022high}. HRRR is a NOAA real-time NWP system over CONUS, updated hourly at 3-km resolution, and is a key operational system for short-term weather prediction. Following common practice in MLWP, we also include two widely used global operational models, the Integrated Forecasting System (IFS) in its high-resolution (HRES) configuration from the European Centre for Medium-Range Weather Forecasts (ECMWF)~\cite{ecmwf2024ifs_cy49r1_obs} and the Global Forecast System (GFS) from NCEP~\cite{noaa2024gfs_bdp_pds}. We evaluate lead times of 1–18 h for near-surface variables and 1–12 h for precipitation. All baselines are interpolated to the native \ours{} grid or to station locations, as appropriate.

\begin{figure}[htbp]
    \centering
\includegraphics[width=1\linewidth]{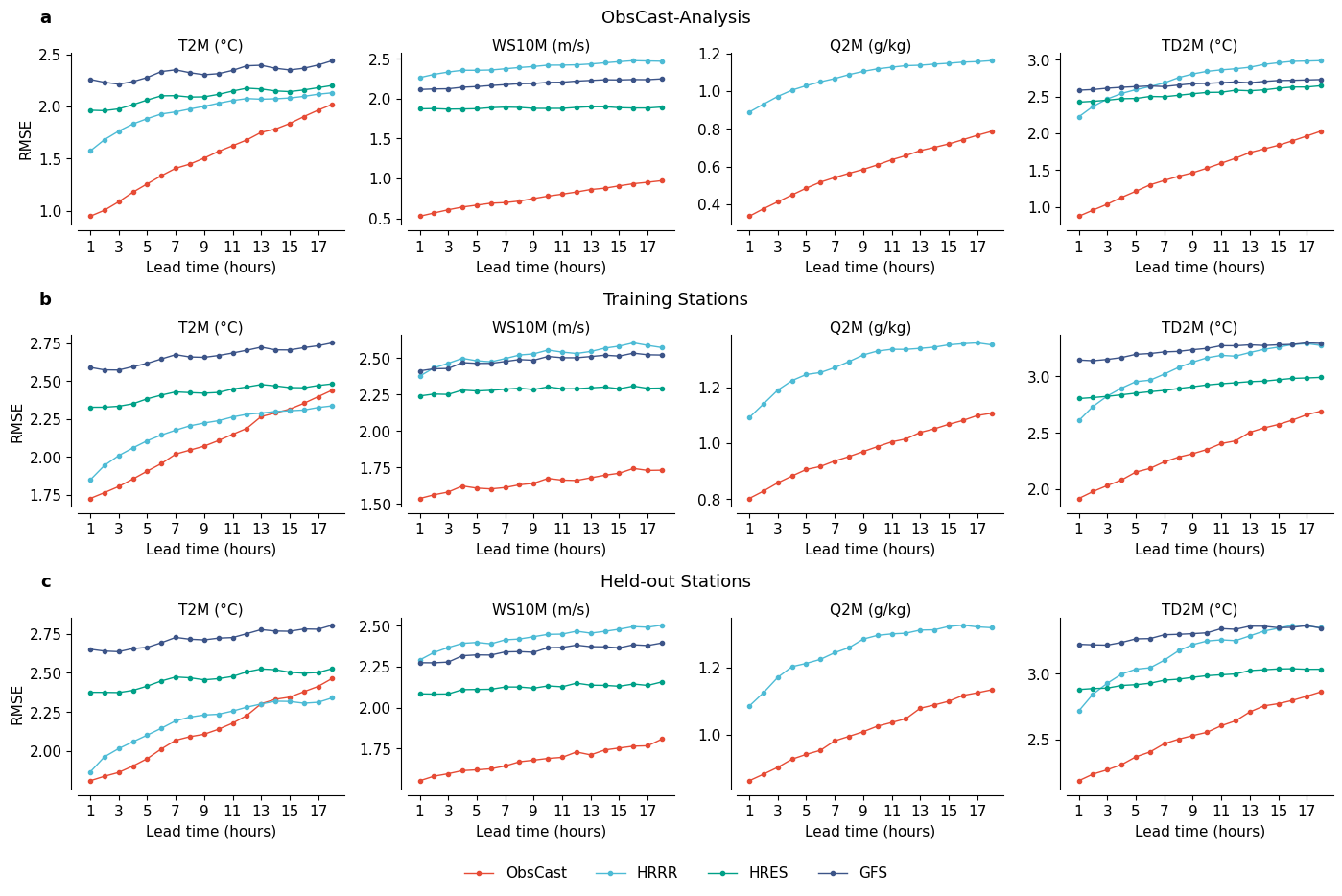}
    \caption{\textbf{Forecast skill of \ours{} over CONUS.} RMSE of \ours{}-Forecast and baseline models as a function of lead time, computed over land grid points or surface station locations. Rows correspond to different verification references: \ours{}-Analysis (\textbf{a}), surface station observations at training locations (\textbf{b}), and surface station observations at held-out locations (\textbf{c}). Columns correspond to different near-surface variables.
}
    \label{fig:fcst-metrics}
\end{figure}

We first evaluate \ours{}-Forecast and the three baselines against the \ours{}-Analysis.  Figure~\ref{fig:fcst-metrics}a shows the CONUS-averaged RMSE for four near-surface variables: 2-m temperature, 10-m wind speed, 2-m specific humidity, and 2-m dew point temperature, as a function of lead time. Across all variables and lead times, \ours{} attains the lowest RMSE, indicating consistently higher forecast skill than HRRR, HRES and GFS, with particularly strong performance for wind speed and humidity. For 2-m specific humidity, comparisons are shown only against HRRR due to data availability in our evaluation dataset. We also find that the margin of improvement narrows with increasing lead time, especially for temperature. This suggests that sustaining the early-lead advantage will require improved representation of boundary forcing and of the three-dimensional atmospheric state evolution beyond the initial condition.

To independently assess the forecast skill of \ours{}, we report RMSE against station observations for both training stations and held-out stations (Figure~\ref{fig:fcst-metrics}b-c). The overall conclusions are consistent with the gridded evaluation against \ours{}-Analysis. Across most variables and lead times, \ours{} achieves lower RMSE than HRRR, HRES and GFS. One exception is for 2-m temperature at the longest lead times, where \ours{} slightly underperforms HRRR beyond 14 h. Absolute RMSE values are larger for all models in the station-based evaluation than in Figure~\ref{fig:fcst-metrics}a, although the relative ordering of models remains broadly unchanged. This increase is expected because point measurements are more affected by localized surface conditions and transient fluctuations that are smoothed in gridded fields. Overall, these station-based results support the main finding that \ours{}-Forecast remains competitive at longer lead times using only raw observations as input, and that it generalizes well to locations that were not seen during training.

In terms of the spatial distribution of RMSE against all training stations, for temperature and dew point, both \ours{}-Forecast and HRRR exhibit larger errors over the western United States, consistent with the influence of complex terrain on near-surface thermodynamics (Extended Data Figure~\ref{fig:fcst-case}a-b). A clear difference emerges along the northern edge of CONUS, where \ours{}-Forecast shows substantially higher RMSE than HRRR at longer lead times. This pattern is consistent with the absence of continuous lateral boundary forcing in \ours{}-Forecast, which limits its ability to represent large-scale influences entering the domain from Canada. In the domain interior and across much of the central and eastern United States, \ours{}-Forecast remains comparable to, or better than, HRRR. \ours{} outperforms HRRR in near-surface moisture forecasting, maintaining lower RMSE for dew point and specific humidity throughout the 18 h forecast window (Extended Data Figure~\ref{fig:fcst-case}b-c). The largest errors are concentrated near the Gulf Coast and the southeastern United States, where moisture gradients and mesoscale variability are strongest, though \ours{} shows a distinct advantage in this region as well as over the western mountain ranges. For 10-m wind speed (Extended Data Figure~\ref{fig:fcst-case}d), a contrasting spatial pattern is observed, where \ours{} exhibits higher RMSE in the central CONUS, while HRRR shows larger RMSE in the western and eastern United States. This behavior is consistent with the apparent positive bias in HRRR relative to station observations reported in earlier studies~\cite{fovell2022evaluation, james2022high}, which inflates RMSE preferentially in regions with lower mean wind speeds. \ours{} learns directly from data, avoiding such large-scale systematic bias and demonstrating the advantages of fully NWP-independent data-driven models over traditional methods.

\begin{figure}[htbp]
    \centering
\includegraphics[width=1\linewidth]{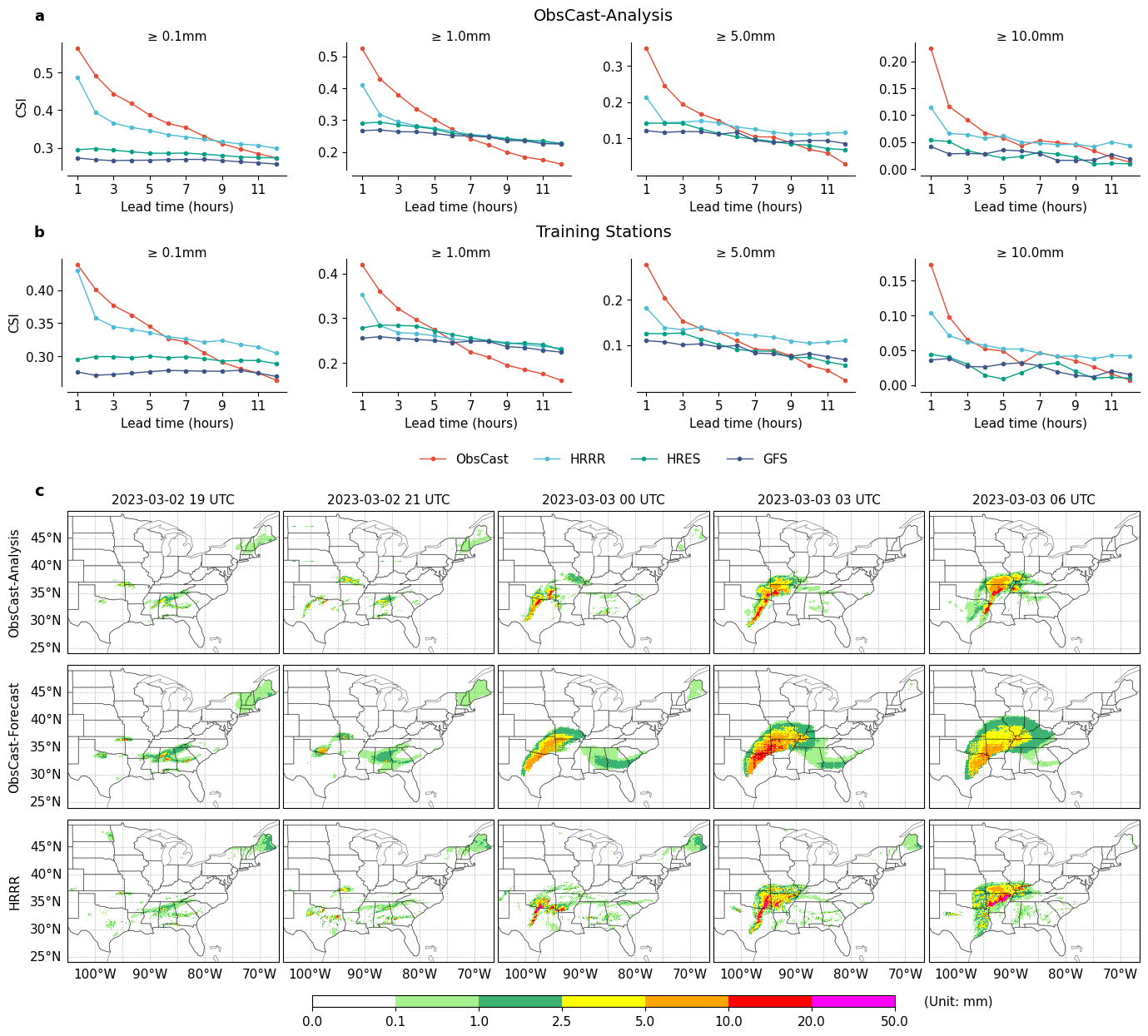}
    \caption{\textbf{Precipitation forecast skill of \ours{} over CONUS.} \textbf{a},\textbf{b}, TS of hourly accumulated precipitation forecasts as a function of lead time at multiple intensity thresholds. Verification is performed against \ours{}-Analysis (\textbf{a}) and against surface station observations (\textbf{b}), comparing \ours{}-Forecast with HRRR and other baselines. \textbf{c}, Example of hourly accumulated precipitation for a representative event, showing the \ours{}-Analysis reference and forecasts from \ours{}-Forecast and HRRR at lead times of 1, 3, 6, 9, and 12 h; both models are initialized at 18:00 UTC on 2 March 2023.
}
    \label{fig:fcst-tp}
\end{figure}

Figure~\ref{fig:fcst-tp}a-b presents hourly precipitation forecast skill using TS at 0.1, 1, 5, and 10 mm/h thresholds against \ours{}-Analysis and station observations. \ours{} significantly outperforms HRRR and other baselines across all thresholds during the initial forecast hours. This superiority persists for 4 to 8 hours, extending beyond traditional nowcasting ranges~\cite{ravuri2021skilful, zhang2023skilful} and indicating the model effectively leverages the richer initial state to capture precipitation dynamics. However, unlike near-surface variables, the skill gap narrows notably after this period. This likely reflects the strong dependence of precipitation on vertical stability and moisture transport. Since these governing variables are only partially observed and represented implicitly, reconstructing mesoscale outcomes becomes increasingly difficult with lead time. Furthermore, higher skill at lower thresholds implies the model is well-calibrated for light precipitation, while heavier rainfall offers room for optimization.

Figure~\ref{fig:fcst-tp}c shows a representative event for hourly accumulated precipitation, comparing the reference \ours{}-Analysis with forecasts from \ours{}-Forecast and HRRR at lead times of 1, 3, 6, 9, and 12 h. In the analysis, precipitation is weak and spatially fragmented at the earliest leads, but a coherent rain band develops rapidly over the south-central United States by 6 to 12 h, with localized heavy cores embedded within the broader system. Despite the limited early precipitation signal, \ours{}-Forecast anticipates both the emergence and subsequent intensification of the event in the correct corridor, reproducing the transition from scattered rainfall to an organized band. This behavior indicates that \ours{}-Forecast is not simply extrapolating existing rain features forward, but can forecast event formation from the evolving initial state inferred from observations.

\subsection{Evaluation of \ours{} over Europe}

\ours{} is a framework that does not rely on region-specific assumptions. To demonstrate its adaptability, we apply the same modeling and training procedure to Europe using region-specific observational inputs. The European setup differs primarily in the geostationary satellite and radar products (see~\nameref{sec:methods}), while the target variables, spatial resolution, and forecast horizon are unchanged. For this region, HRES and GFS are used as baselines for evaluating the performance of \ours{}.

For \ours{}-Analysis, we use HRES T0 as the baseline, where T0 denotes the zero-hour forecast. Similar to Figure~\ref{fig:anl-metrics}a, \ours{} yields consistently lower RMSE than HRES T0 for both training and held-out stations (Extended Data Figure~\ref{fig:anl-metrics-eu}). Relative to HRES T0, RMSE is reduced by 30.0\%, 31.8\%, 37.1\%, and 35.0\% for 2-m temperature, 2-m dew-point temperature, 10-m wind speed, and 2-m specific humidity on training stations, and by 29.8\%, 35.8\%, 33.6\%, and 38.3\% on held-out stations for the same variables. The comparable improvements on held-out stations indicate that the model generalizes well across Europe. The larger margins, compared with the CONUS results against RTMA, likely reflect differences in the baselines, as HRES has coarser native resolution and exhibits weaker consistency with station observations than RTMA.

For hourly precipitation (Extended Data Figure~\ref{fig:anl-metrics-eu}b), \ours{}-Analysis achieves higher TS than HRES T0 at all evaluated thresholds for both training and held-out stations, with the largest gains at moderate and heavy precipitation. Consistent improvements are also observed in the POD and FAR. As the threshold increases, the number of qualifying records decreases. Particularly, at the 10 mm/h threshold for held-out stations, HRES T0 produces no exceedances, rendering FAR undefined.

\begin{figure}[htbp]
    \centering
\includegraphics[width=0.9\linewidth]{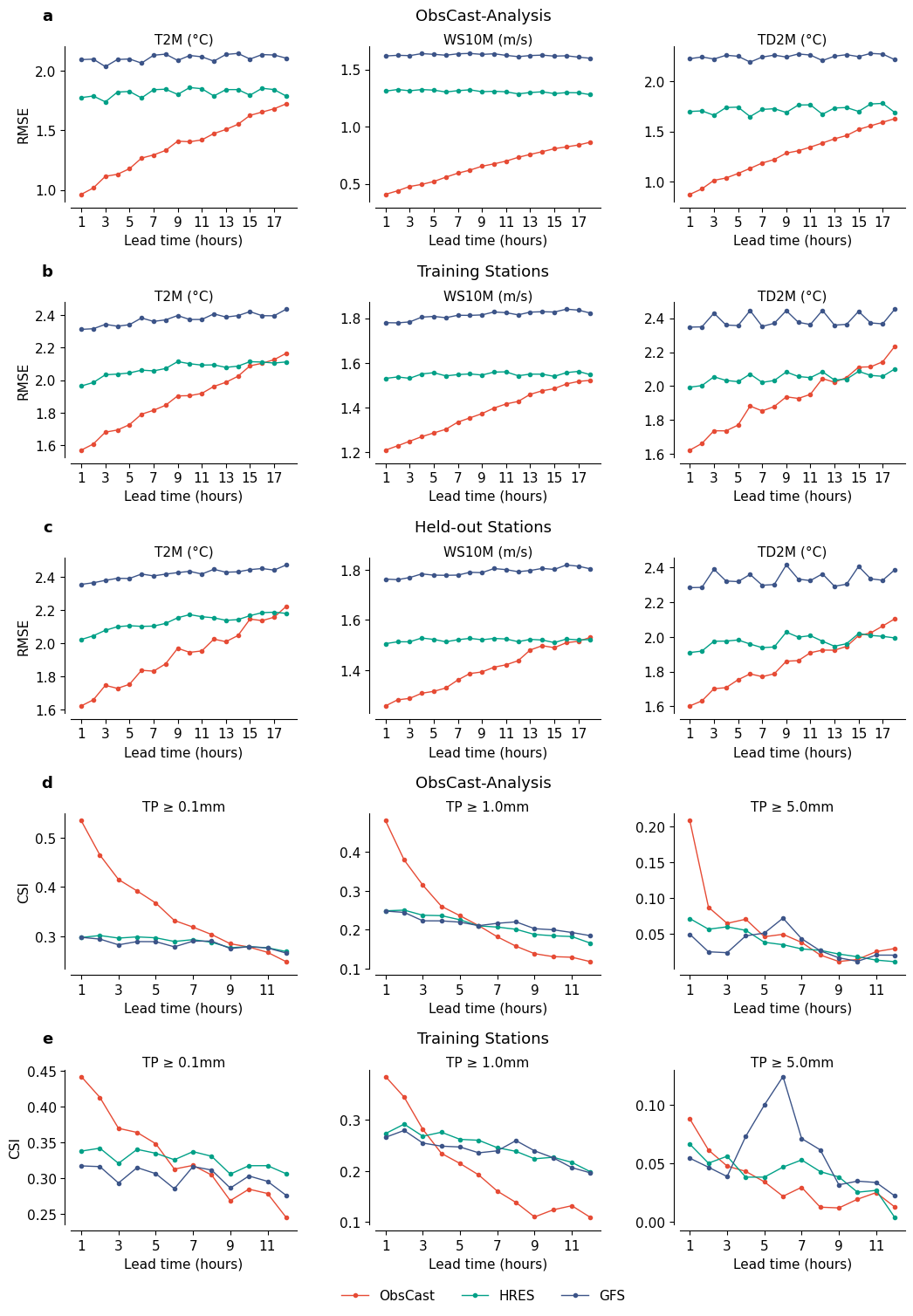}
    \caption{\textbf{Forecast skill of \ours{} over Europe.} Performance of \ours{} compared with baseline models as a function of lead time. (\textbf{a}--\textbf{c}) RMSE of near-surface variables evaluated over land grid points or surface station locations, with verification against \ours{}-Analysis (\textbf{a}), surface station observations at training locations (\textbf{b}), and surface station observations at held-out locations (\textbf{c}). (\textbf{d}--\textbf{e}) TS of hourly accumulated precipitation forecasts at multiple intensity thresholds, verified against \ours{}-Analysis (\textbf{d}) and surface station observations (\textbf{e}).
}
    \label{fig:fcst-eu}
\end{figure}

The forecast skill of \ours{}-Forecast over Europe is evaluated against three references: \ours{}-Analysis, training-station observations, and held-out station observations. Figure~\ref{fig:fcst-eu}a-c reports domain-averaged RMSE as a function of lead time for 2-m temperature, 10-m wind speed, and 2-m dew-point temperature. Consistent with the CONUS results, \ours{}-Forecast exhibits lower RMSE than HRES and GFS throughout the 1–18 h range in most cases, with the only exception being that 2-m temperature and dew-point temperature verified against station observations are slightly worse than HRES in the final forecast hours. Across variables, the skill gap gradually narrows with lead time, for reasons consistent with those discussed in section~\ref{results}. Overall, the European configuration preserves the main advantage of \ours{}-Forecast and shows comparable performance on held-out stations, supporting generalization to locations not used during training.

Figure~\ref{fig:fcst-eu}d-e presents hourly precipitation skill using TS at the 0.1, 1.0, and 5.0 mm/h threshold. The 10.0 mm/h threshold is excluded here due to insufficient sample size within the European test set. When verified against \ours{}-Analysis, \ours{}-Forecast achieves higher TS than HRES at the 0.1 mm/h threshold through the first 9 hours, with the largest gains concentrated at the earliest lead times. Verification against station observations yields a shorter-lived advantage, with \ours{}-Forecast exceeding HRES for roughly the first 5 hours. At the 5.0 mm/h threshold, the duration of this advantage is shortest. A likely contributor is the limited availability of precipitation station reports in Europe (Figure~\ref{fig:num_stns}), which reduces direct supervision and yields fewer qualifying samples. This suggests that additional improvements may be needed to better handle precipitation forecasting in data-sparse regimes.

\section{Discussion}\label{discussion}

We introduce \ours{}, a regional, high-resolution weather forecasting system that is completely independent of NWP from training to inference. The system ingests heterogeneous raw observations from satellites, radars, stations, and other sources to produce hourly gridded analyses and forecasts over target regions. The noisy and sparse nature of raw observations poses significant challenges compared to relying solely on NWP-based reanalyses~\cite{janjic2018representation}. Nevertheless, across evaluations over CONUS and Europe, \ours{} yields analysis fields that are more consistent with independent observations than existing analysis products, while its forecasts achieve skill comparable to or exceeding that of mainstream operational systems over the first 18 hours. A key design choice relative to prior end-to-end approaches is the separation into two coupled but operationally separate components: an analysis model that maps observations to dense gridded fields using only sparse station-based supervision, and a forecast model that learns atmospheric evolution in that grid space. This design avoids reliance on NWP-based reanalyses common to most end-to-end models, while closing the performance gap that has previously constrained approaches trained directly on sparse observations.

While \ours{} already produces forecasts for multiple variables of greatest practical interest, its explicit outputs are presently limited to near-surface quantities, with upper-air information only implicitly encoded in the latent state. Extending the system to explicitly model the three-dimensional atmosphere is a natural next step and appears feasible in light of the present results. It in turn motivates integrating additional observational sources to constrain the vertical structure. As a limited-area model, forecasts are generally strongly influenced by large-scale circulation outside the target domain~\cite{warner1997tutorial}. \ours{} partially addresses this dependence by incorporating observations over a broader spatial extent when forming the initial state. Further improvements at longer lead times may benefit from incorporating time-evolving large-scale guidance, such as global model forecasts, to better capture boundary influences~\cite{xu2025artificial}. Finally, deeper interpretability study remains an important direction, including quantifying the contributions of individual observing platforms to specific weather processes. Such insight could also support more efficient data utilization, particularly in regions where observational coverage is sparse. These limitations point to broader machine-learning challenges in observation-driven forecasting.

The high forecast skill of \ours{} represents a substantive step toward purely data-driven weather forecasting systems, a paradigm with significant practical implications for weather services. It provides evidence that machine learning can learn the governing weather structure directly from observations without being explicitly constrained by hand-crafted physical priors, thereby fully realizing an end-to-end approach. Moreover, the analysis component is trained against in-situ measurements, anchoring the learned gridded states more directly to surface conditions experienced by end users and reducing reliance on operational post-processing and bias correction~\cite{dueben2018challenges}. Because MLWP can be substantially faster, less computationally intensive, and more straightforward to improve than conventional pipelines, the approach also offers flexibility to tailor domains and variables to specific needs, including regional extreme weather prediction and downstream applications for agriculture and renewable energy. More broadly, organizations with sufficient regional data could incorporate local observations to improve performance on their target domain without the substantial burden of building and operating NWP systems.

\section{Methods}\label{sec:methods}

\subsection{Data Sources}

\textbf{Study domains and target grids.} We conducted experiments over CONUS and Europe, and used a larger context region for the forecast model in each case. All input data and model labels were interpolated onto regular $0.05^\circ \times 0.05^\circ$ latitude--longitude grids. For CONUS, the target region spans $24.70^\circ$ to $50.30^\circ$ latitude and $-128.00^\circ$ to $-64.00^\circ$ longitude, and the corresponding larger context region spans $10.00^\circ$ to $69.75^\circ$ latitude and $-160.00^\circ$ to $-30.25^\circ$ longitude. For Europe, the target region spans $36.10^\circ$ to $56.10^\circ$ latitude and $-10.70^\circ$ to $25.25^\circ$ longitude, and the corresponding larger context region spans $21.10^\circ$ to $71.05^\circ$ latitude and $-35.00^\circ$ to $54.80^\circ$ longitude.

\medskip\noindent\textbf{Ground station observations.} To accurately estimate near-surface atmospheric conditions and precipitation, ground station observations are the primary data source, serving as inputs to both \ours{}-Analysis and \ours{}-Forecast, and as labels for \ours{}-Analysis. These data come from WeatherReal-Synoptic~\cite{jin2024weatherreal}, which aggregates observations from hundreds of public and private station networks worldwide, including operational SYNOP and METAR observations within the WMO global data exchange framework. WeatherReal applies standard quality control algorithms to detect and remove suspect station reports. Station observations used in this work as labels include 2-m temperature, 2-m dew point temperature, 2-m specific humidity, 10-m wind speed, and hourly precipitation accumulation. The station inputs comprise 10 channels in total, including five additional observed or derived quantities. The number of stations increased steadily from 2018 to 2024 (Extended Data Figure~\ref{fig:num_stns}). Observations of 2 m air temperature were collected from the largest number of stations, increasing from approximately 16,000 to 23,000 over this period in CONUS. In contrast, precipitation was measured at substantially fewer stations, particularly during the early years in Europe, with only about 500 stations available.

\medskip\noindent\textbf{Remote sensing data for \ours{}-Analysis over CONUS.} Because ground stations are spatially sparse and primarily sample near surface conditions, we additionally ingest remote sensing observations to provide continuous spatial context. For \ours{}-Analysis over CONUS, the primary remote sensing input is multispectral imagery from the Geostationary Operational Environmental Satellite 16 (GOES 16)~\cite{GOES16_ABI}. Its Advanced Baseline Imager provides 16 spectral bands spanning the visible, near infrared, and infrared, with rapid refresh over the CONUS sector. All 16 channels are used as input to capture the distribution of clouds, atmospheric moisture, and thermal emission that inform near surface conditions. To ensure that \ours{} inputs remain free of any processing related to NWP, only low-level satellite observations are used. In \ours{}-Analysis, precipitation is estimated by an independently trained model that incorporates weather radar observations as an additional constraint. Over CONUS, we use the Seamless Hybrid Scan Reflectivity and Radar Precipitation Rate products from the Multi Radar Multi Sensor (MRMS) system~\cite{zhang2016multi}, which supplements the spatiotemporal coverage not provided by station observations and provides a consistent proxy for precipitation related atmospheric structure.

\medskip\noindent\textbf{Remote sensing data for \ours{}-Analysis over Europe.} Over Europe, where GOES-16 and MRMS coverage is not available, we switch to geostationary satellite imagery from Meteosat\cite{EUMETSAT_Meteosat} and weather radar products from the EUMETNET OPERA~\cite{saltikoff2019opera}. Meteosat SEVIRI imagery provides multispectral visible and infrared observations over Europe. We use the 11 standard SEVIRI channels. OPERA aggregates radar observations from national networks to produce harmonized radar reflectivity products.

\medskip\noindent\textbf{Additional context for \ours{}-Forecast.} To provide broader scale context that influences limited area evolution but is not covered by regional observations, the forecast model ingests additional upper air observations over a larger domain that spans twice the spatial extent of the target region. Specifically, we include polar orbiting microwave sounder products comprising AMSU-A~\cite{weng2003advanced} and MHS~\cite{10.1117/12.737986}, flown on NOAA and MetOp satellites. AMSU-A contributes temperature sounding information from 15 channels, with data separated into four groups for each channel, defined by platform (NOAA and MetOp) and orbit direction (ascending and descending). MHS contributes humidity sounding information from 5 channels with the same grouping. In addition, we include radiosonde vertical soundings from the Integrated Global Radiosonde Archive (IGRA)~\cite{Durre2016IGRAv2}, represented as sparse profiles with 78 channels. It includes geopotential, temperature, U- and V-components of wind, relative humidity, and specific humidity at 13 pressure levels (1000, 925, 850, 700, 600, 500, 400, 300, 250, 200, 150, 100, 50 hPa).

\medskip\noindent\textbf{Static features and time encodings.} To account for diurnal and seasonal variability, we augment the model inputs with time encodings represented using sinusoidal functions, with two components derived from the hour of day and two components derived from the month of year. We additionally include constant topographic variables to help the model distinguish surface characteristics across regions. The topographic fields are derived from the Global Multi resolution Terrain Elevation Data 2010~\cite{danielson2011global}.

\subsection{Data Pre-processing}

\textbf{Temporal and spatial alignment.} Station and remote-sensing observations are mapped to a regular latitude-longitude grid at 0.05° resolution prior to ingestion by the model. Gridded remote-sensing data are interpolated to the model grid using bilinear interpolation. Station data are mapped to the nearest grid point.

\medskip\noindent\textbf{Normalization and missing data handling.} Each input variable is normalized to the range of [-1, 1]. Missing values in the gridded inputs are filled with zero after normalization and interpolation.

\medskip\noindent\textbf{Station data partitioning and utilization.} Approximately 800 stations in CONUS and 280 stations in Europe were randomly selected as held-out datasets, excluded from both model inputs and supervision during training of all components, and used exclusively for independent evaluation. During training, supervision in \ours{}-Analysis is applied only at grid locations with available station observations. At each labeled grid cell, the corresponding observation is masked from the input, and the model is trained to reconstruct it from the surrounding spatiotemporal context. Within the local 25 × 25 grid kernel, station observations are additionally removed at random using multiple masking strategies to improve robustness to spatial sparsity and heterogeneous reporting patterns.

\medskip\noindent\textbf{Land masking.} Because supervision for \ours{}-Analysis is derived exclusively from surface station observations, which are located over land, all model training is restricted to land grid points. A static land–sea mask is applied to both inputs and outputs, and losses are computed only over land grid cells. All reported analyses, forecasts, and evaluation metrics in this work are therefore restricted to land areas.

\begin{table*}[t]
\centering
\footnotesize
\setlength{\tabcolsep}{5pt}
\renewcommand{\arraystretch}{1.15}

\caption{Summary of data sources used in the end-to-end regional AI weather forecasting system.}
\label{tab:data_sources}

\newcolumntype{Y}{>{\raggedright\arraybackslash}X}

\begin{tabularx}{\textwidth}{@{} l Y c Y @{}}
\toprule
\textbf{Data Category} & \textbf{Dataset} & \textbf{Number  of Variables} & \textbf{Role in Framework} \\
\midrule

Ground-based observations
& WeatherReal-Synoptic
& 10
& Inputs and labels for Analysis; inputs for Forecast. \\
\cmidrule(lr){1-4}

Weather radar
& MRMS (CONUS); OPERA (Europe)
& 2
& Inputs for Analysis and Forecast. \\
\cmidrule(lr){1-4}

Geostationary satellite
& GOES-16 Level 1b (CONUS); Meteosat SEVIRI Level 1.5 (Europe)
& 16 (CONUS); 11 (Europe)
& Inputs for Analysis and Forecast. \\
\cmidrule(lr){1-4}

\multirow[c]{2}{*}{Polar-orbiting satellite}
& NOAA/MetOp AMSU-A Level 1b
& $15 \times 4$
& \multirow[c]{2}{*}{Inputs for Forecast.} \\
& NOAA/MetOp MHS Level 1b
& $5 \times 4$
& \\
\cmidrule(lr){1-4}

Radiosonde observations
& IGRA
& 78
& Inputs for Forecast. \\
\cmidrule(lr){1-4}

Static and temporal forcings
& Time and topography
& 33
& Inputs for Analysis and Forecast. \\

\bottomrule
\end{tabularx}
\end{table*}

\subsection{Temporal splits for training, validation, and testing}

\textbf{CONUS model.} For the \ours{} model trained over CONUS, the training set spans 2018–2022, while all evaluation data are drawn from 2023 using fixed day-of-month partitions. The validation set comprises the 6th–8th day of each month, and the test set comprises the 1st–3rd day of each month.

\medskip\noindent\textbf{Europe model.} For the \ours{} model trained over Europe, the training set spans 2018–2023, and evaluation is conducted analogously on 2024, with validation taken from the 6th–8th day of each month and testing from the 1st–3rd day of each month.

\subsection{The \ours{} Model}

\textbf{Model architecture of \ours{}-Analysis.} \ours{}-Analysis reconstructs gridded analysis fields by predicting the value at each grid point from local spatiotemporal context. For a target grid cell at time $t$, the model ingests an observation array of shape $C \times T \times H \times W$ extracted from a window centered on that cell, where $C$ is the number of observational channels, $T$ is the number of temporal frames, and $H \times W$ denotes the number of latitude and longitude points in the spatial window. We set $H=W=25$, corresponding to a $1.25^{\circ} \times 1.25^{\circ}$ neighborhood at the native $0.05^{\circ}$ resolution. Temporal context is provided with $T=3$ frames spanning $t-1$\,h to $t+1$\,h. For computational efficiency, all target locations at a given time step are batched so that the corresponding outputs are produced in parallel.

The analysis network follows a Vision Transformer (ViT) architecture~\cite{dosovitskiy2020image}. Using a patch size of 1, each grid cell is treated as a token. The channel and time dimensions ($C$ and $T$) are flattened for each sample and linearly projected into a $D$-dimensional embedding, yielding a representation of size $H \times W \times D$. This representation is processed by 8 ViT blocks to capture spatiotemporal interactions within the local window. A final linear projection maps the transformer output back to physical space, producing an analysis array of size $H \times W \times O$, where $O$ denotes the number of output variables. The near-surface analysis jointly predicts $O=4$ variables, whereas precipitation is produced by a separate model of the same architecture with $O=1$.

\medskip\noindent\textbf{Model architecture of \ours{}-Forecast.} \ours{}-Forecast consists of two parts: the Conditioning Block, which transforms the input observations into an initial state, and the Forecasting Block, which makes the forward predictions. This dual-module architecture allows different architectures to be used to the fullest extent for the two tasks: The Conditioning Block leverages the Swin Transformer (SwinT)~\cite{liu2022swintransformerv2scaling}, which effectively captures localized features with its attention mechanism. This makes it an excellent choice for purposes of data completion and feature extraction. The Forecasting Block incorporates the adaptive Fourier neural operators~\cite{guibas2021adaptive}, which uses Fourier transforms to efficiently model meteorological dependencies across different scales.

The Conditioning Block is designed to convert heterogeneous raw observations into a structured initial-state representation. The block begins by encoding each observational stream with a dedicated SwinT-based encoder. All regional inputs over the target domain are processed with a shared regional encoder. For the large-scale context inputs, we use separate encoders to reflect differences in their physical meanings. In particular, each acquisition group in the two polar orbit satellite data sources is processed by its own encoder. Each encoder begins with a 3D patch embedding layer to form a two-dimensional grid of tokens. This layer is followed by a feature extraction process consisting of SwinT blocks. The resulting representations are then fused across data sources using cross-attention. The regional encoder output provides the query embeddings, while the background encoders provide key–value embeddings, allowing the model to selectively extract large-scale information most relevant to the regional state. Finally, the cross-attended representation is processed by a decoder, also implemented with SwinT blocks, to produce the initial-state representation. By leveraging the shifted-window attention mechanism, raw observations are gathered to form a robust and accurate estimate of the current conditions.

The output of the Conditioning Block includes two parts. The first part consists of channels corresponding to the target variables at the initial time step. These channels are supervised against the labels, so that the model learns the mapping from sparse observations to dense representations. The second part consists of 36 channels, which do not participate in the loss computation. Instead, these channels are designed to encode and propagate implicit information, which may include upper-air dynamics and lateral boundary conditions. Together, these channels serve as input to the Forecasting Block, allowing the model to generate a complete meteorological representation to use for prediction.

The Forecasting Block uses the learned initial state to generate autoregressive hidden states in a latent space. The resulting sequence of patch embeddings is then processed by a stack of adaptive Fourier neural operator (AFNO) layers~\cite{guibas2021adaptive}, where the time and feature dimensions are combined to enable efficient spatiotemporal mixing.

At each autoregressive step, the hidden state is then mapped to physical variables through two AFNO-based decoders. One decoder predicts the near-surface variables together with the geostationary satellite channels, while a second decoder is dedicated to precipitation prediction. A linear projection is applied to recover gridded fields at the native resolution.

The forecast model is conditioned on the observations within the last 6 h, spanning $t_0-5$ to $t_0$, to construct the initial state. Forecasts are then produced autoregressively in three consecutive 6-hour segments: the first rollout predicts hours $t_0+1$ to $t_0+6$, the second predicts $t_0+7$ to $t_0+12$, and the third predicts $t_0+13$ to $t_0+18$.

\subsection{Training Methods}

\textbf{Training procedures.} All components of \ours{} were optimized with Adam, using a learning-rate schedule with linear warmup followed by linear decay. To reduce overfitting, training was stopped early on the basis of validation performance. The number of optimization steps and batch sizes varied across \ours{} components.

\medskip\noindent\textbf{Training objective.} Continuous near-surface variables were optimized with a hybrid regression loss defined as an equally weighted sum of mean squared error and mean absolute error. For \ours{}-Analysis, this loss was computed only at labeled land grid cells collocated with available station observations. For \ours{}-Forecast, the same regression loss was computed jointly over land grid points, surface-variable channels, and supervised time steps. Supervision spanned four temporal windows: the conditioning interval \(t_0-5\) to \(t_0\), and three autoregressive forecast intervals, \(t_0+1\) to \(t_0+6\), \(t_0+7\) to \(t_0+12\), and \(t_0+13\) to \(t_0+18\). Loss contributions from all supervised time steps were averaged uniformly, with no additional weighting across the temporal windows. \ours{}-Forecast was also trained to predict the geostationary satellite channels at each of the 18 forecast lead times, with the corresponding regression loss included in the training objective as an auxiliary loss.

Precipitation was modeled as a categorical target rather than as a direct regression output. Hourly accumulated precipitation was discretized into nine bins: 0--0.1, 0.1--1.0, 1.0--2.5, 2.5--5.0, 5.0--10.0, 10.0--20.0, 20.0--50.0, 50.0--100.0, and \(>100.0\) mm. The precipitation objective combined a weighted multiclass cross-entropy loss over these bins with an additional binary classification loss for rain occurrence, to account for class imbalance and improve discrimination between raining and non-raining conditions.

Training of \ours{}-Forecast used two separately trained models with identical architecture but different loss configurations. In the first training run, the model was optimized using only the loss for the near-surface variables, while the precipitation loss was disabled. The resulting model was used to generate forecasts for the near-surface variables. In the second training run, the same architecture was again initialized from scratch, but all loss terms were activated, including the precipitation objective. The resulting model was used for precipitation forecasting. In this second model, supervision of the near-surface variables served to refine the shared representation, although only the precipitation output was retained in the final results.

\section{Data availability}

Most of the datasets used in this study are publicly available from the providers listed below. Radar data were obtained from the Multi-Radar/Multi-Sensor (MRMS) system (\url{https://www.nssl.noaa.gov/projects/mrms/}) and the European OPERA radar data service via EUMETNET MeteoGate (\url{https://radar.meteogate.eu/}). Geostationary satellite data were obtained from GOES (\url{https://www.goes.noaa.gov/}) and Meteosat via the EUMETSAT Data Store (\url{https://data.eumetsat.int/}). Microwave sounding observations from AMSU-A/MHS were obtained from NOAA CLASS for NOAA platforms (\url{https://www.class.noaa.gov/}) and from the EUMETSAT Data Store for MetOp platforms (\url{https://data.eumetsat.int/}). Upper-air observations were obtained from the IGRA (\url{https://www.ncei.noaa.gov/products/weather-balloon/integrated-global-radiosonde-archive}), and near-surface observations from the WeatherReal-Synoptic benchmark, distributed through the WeatherReal-Benchmark repository (\url{https://github.com/microsoft/WeatherReal-Benchmark}). Baseline forecast data used for comparison were obtained from IFS-HRES (\url{https://www.ecmwf.int/en/forecasts/datasets/set-i}), HRRR (\url{https://rapidrefresh.noaa.gov/hrrr/}), and GFS (\url{https://www.ncei.noaa.gov/products/weather-climate-models/global-forecast}).

\section{Code availability}
The implementation of \ours{} will be released on GitHub at or before the time of publication.

\bibliography{V0.1}

\section{Acknowledgments}
The authors express their gratitude to all providers for providing the publicly accessible data.
%NOAA for providing publicly accessible RTMA analysis data and GOES satellite data.
The authors would like to thank Synoptic Data PBC (accessible at \url{https://synopticdata.com/}) for aggregating station observations and providing the Mesonet API for us to download those data, which are crucial in building an AI-based framework like \ours{} for weather forecasting.

\section{Contributions}
P.Z., S.X. and H.D. conceived the study and designed the overall methodology. P.Z., S.X., Z.N. and Z.F. curated and processed the data. P.Z., S.X. and W.J. developed the model, conducted the experiments and analyzed the results. W.J. drafted the manuscript, with contributions from J.B. and input from all authors. R.T. and J.W. contributed to scientific framing, interpretation of results and manuscript revision. K.T., H.S., B.Z. and Q.Z. provided scientific supervision and project oversight. H.D. supervised and coordinated the study. All authors reviewed and approved the manuscript.

\section{Competing interests}
The authors declare no competing interests.

\section{Extended data figures}

\begin{extdatafig}[htbp]
    \centering
\includegraphics[width=1\linewidth]{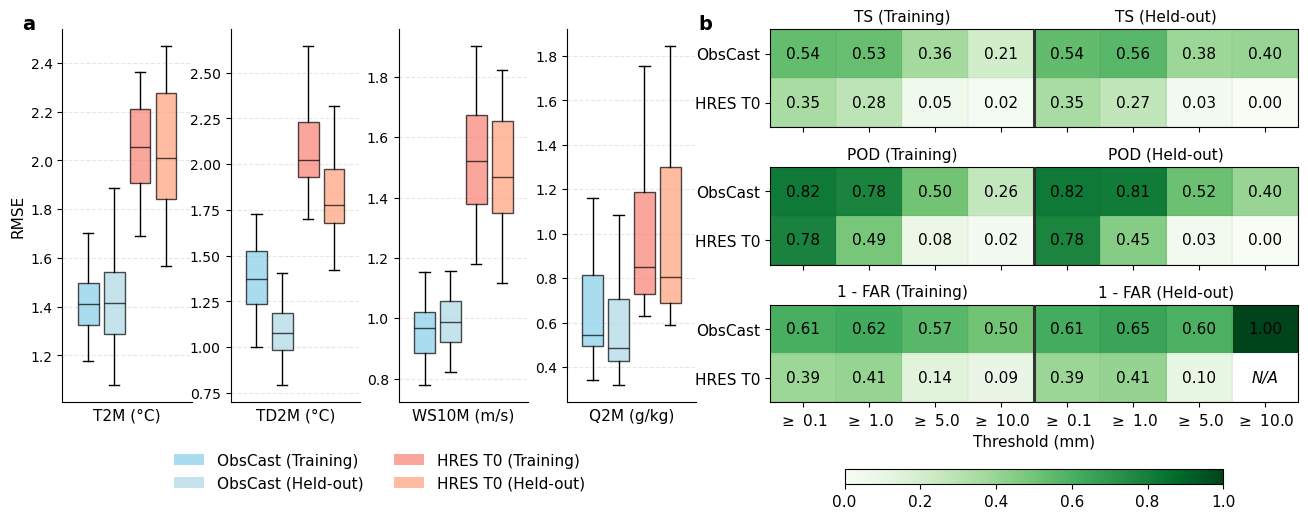}
    \caption{\textbf{Evaluation of \ours{}-Analysis over Europe.} Station-wise RMSE of near-surface variables (\textbf{a}) and skill of hourly accumulated precipitation (\textbf{b}) for \ours{}-Analysis, evaluated against surface station observations. Results are shown separately for stations used during training and for held-out stations, with ECMWF HRES T0 used as the reference baseline.
}
    \label{fig:anl-metrics-eu}
\end{extdatafig}

\begin{extdatafig}[htbp]
    \centering
\includegraphics[width=1\linewidth]{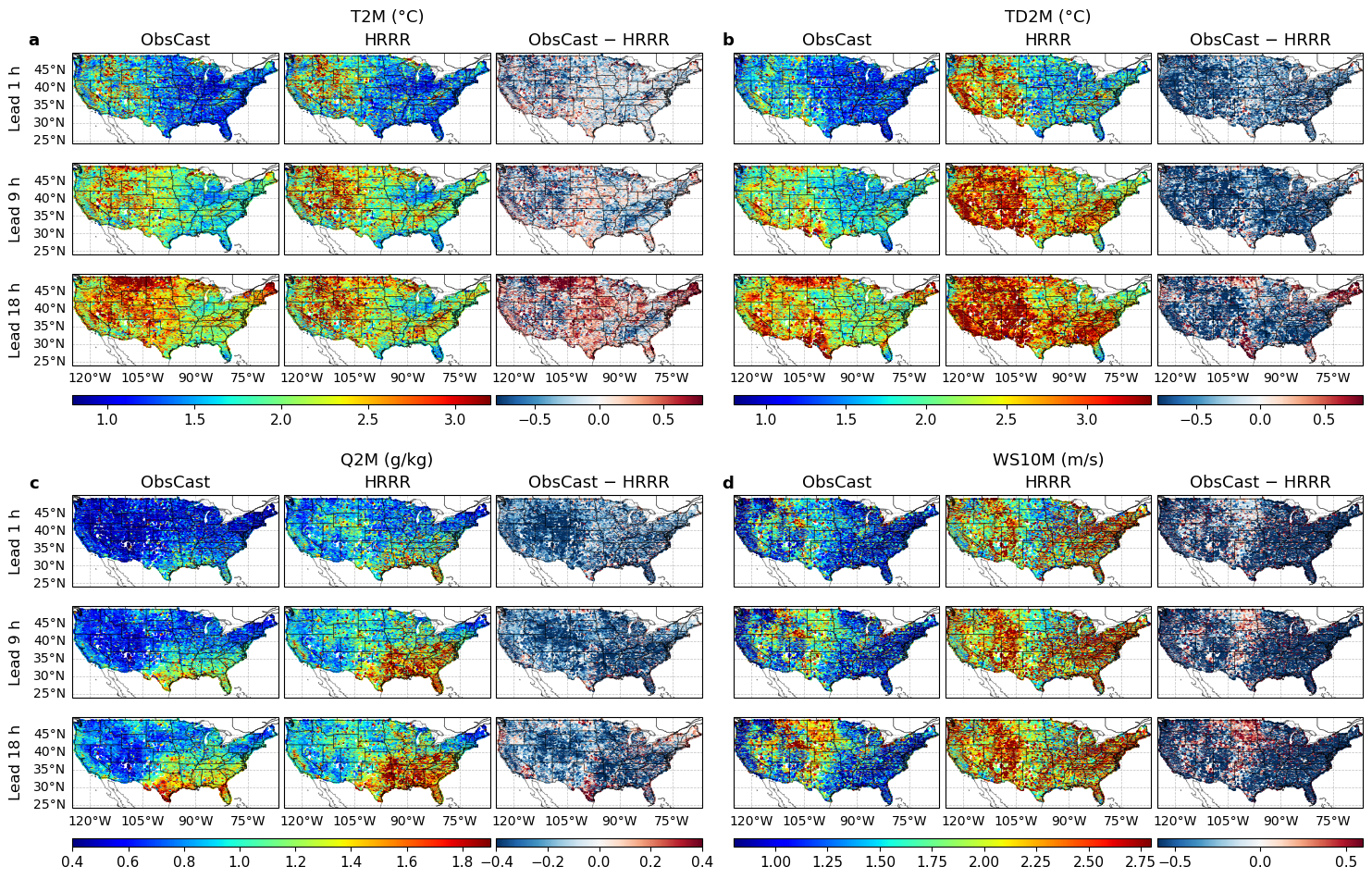}
    \caption{\textbf{Spatial distribution of forecast errors over CONUS.} Spatial maps of RMSE evaluated at surface station locations for four near-surface variables. Each subplot corresponds to a different variable. Columns show RMSE for \ours{}-Forecast, RMSE for HRRR, and their difference (\ours{} minus HRRR), respectively. Rows correspond to forecast lead times of 1, 9, and 18 h.
}
    \label{fig:fcst-case}
\end{extdatafig}

\begin{extdatafig}[htbp]
    \centering
\includegraphics[width=1\linewidth]{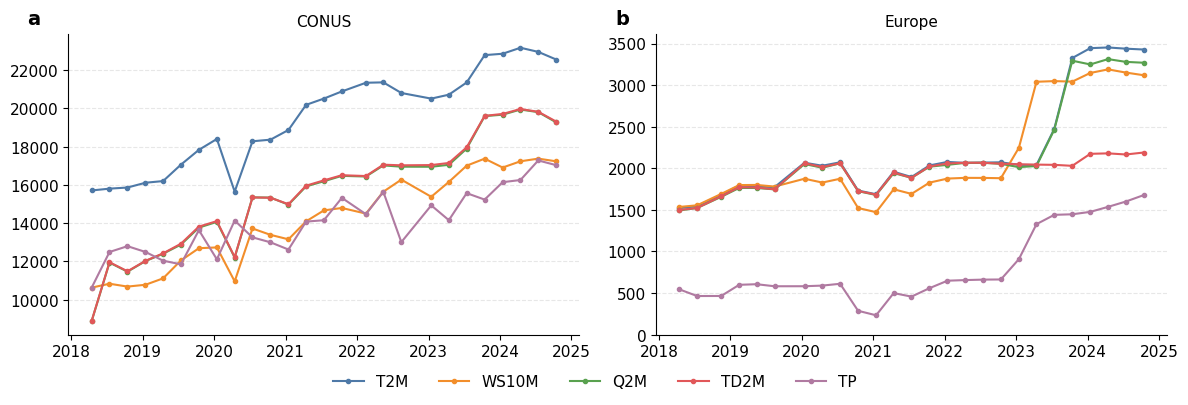}
    \caption{\textbf{Temporal evolution of surface station availability.} Number of reporting surface stations over time for different observed variables used in this study, shown separately for CONUS (\textbf{a}) and Europe (\textbf{b}). Each point represents the average number of stations within a 3-month moving window, plotted at 3-month intervals.
}
    \label{fig:num_stns}
\end{extdatafig}

\clearpage
\section{Supplementary Information}

\setcounter{figure}{0}
\counterwithin{figure}{section}

\subsection{Impact of training labels for \ours{}-Forecast}

\begin{figure}[htbp]
    \centering
\includegraphics[width=1\linewidth]{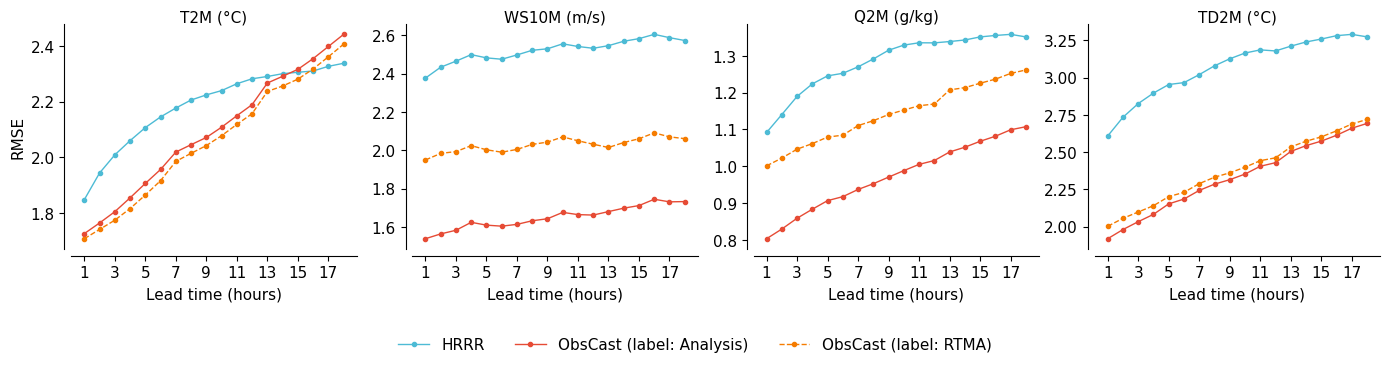}
    \caption{\textbf{Impact of training labels on \ours{}-Forecast skill over CONUS.}
RMSE of \ours{}-Forecast as a function of lead time for four near-surface variables, evaluated at CONUS surface stations, when trained with \ours{}-Analysis versus RTMA as supervision labels. HRRR is shown as a reference.
}
    \label{fig:rtma}
\end{figure}

To assess whether the benefits of \ours{} arise specifically from its learned analysis fields, we performed an ablation in which the forecast model was trained with RTMA rather than \ours{}-Analysis as the supervision target, while keeping the forecast architecture, inputs, and optimization settings unchanged (Figure~\ref{fig:rtma}). This comparison shows that using \ours{}-Analysis as the training label improves downstream forecast quality for most near-surface variables over CONUS. Relative to RTMA-based supervision, the \ours{}-Analysis labels yield markedly lower RMSE for 10-m wind speed and 2-m specific humidity across the full 1–18 h forecast range, while 2-m dew-point temperature remains broadly comparable and 2-m temperature shows only a slight degradation. These results are consistent with the finding that \ours{}-Analysis agrees more closely with station observations than RTMA, indicating that the learned analysis component does not merely replace a conventional gridded product, but provides a more effective target representation for training the forecast model. In this sense, the two-stage design of \ours{} is functionally important. These improved observation-driven analyses translate directly into improved forecast skill, even in a region where a high-resolution operational reanalysis is already available.

\subsection{Impact of station data availability at inference time}

Because surface station observations in operational settings are subject to intermittent outages and reporting delays, it is important to assess how forecast skill degrades when a fraction of station inputs is withheld.
We evaluated the sensitivity of \ours{}-Forecast to reduced station availability at inference time by randomly withholding 30\% and 50\% of station inputs from the test set, while leaving all non-station inputs unchanged. Forecast errors increase only slightly under both perturbations, with the 30\% dropout case showing nearly indistinguishable RMSE from the full-input setting across all variables and lead times, and the 50\% dropout case producing only modest degradation through 18 h. Importantly, even under this severe and operationally unlikely level of station loss, performance remains well within the margin by which \ours{} outperforms HRRR. This result indicates that \ours{}-Forecast does not depend excessively on any particular subset of surface stations, but instead integrates information redundantly across heterogeneous observations. The model is therefore robust to intermittent outages, reporting delays, and spatial irregularities that commonly affect real-time station networks, supporting its practical suitability for operational deployment.

\begin{figure}[htbp]
    \centering
\includegraphics[width=1\linewidth]{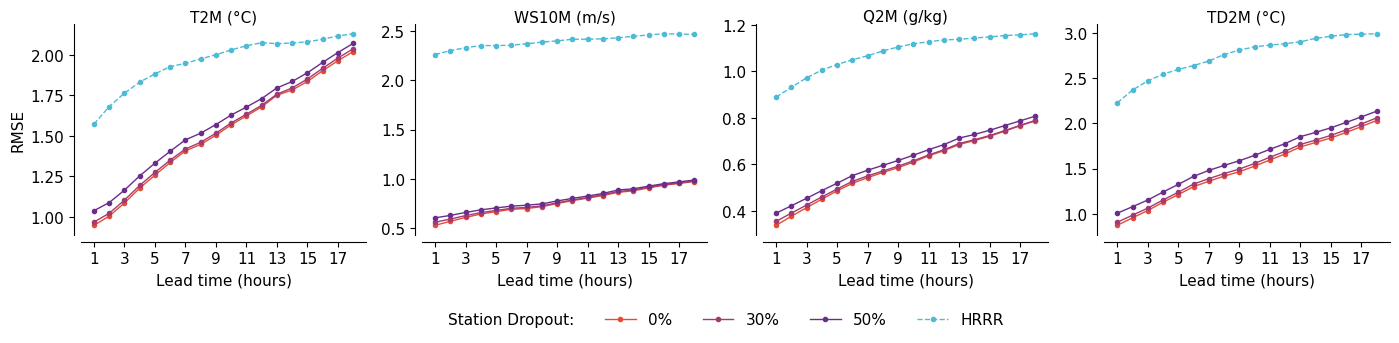}
    \caption{\textbf{Robustness of \ours{}-Forecast to reduced station availability at inference time.}
RMSE of \ours{}-Forecast as a function of lead time for four near-surface variables, verified against \ours{}-Analysis, under three levels of station dropout applied only at inference: 0\% (all stations available), 30\%, and 50\%.
}
    \label{fig:robustness}
\end{figure}

\end{document}